\documentclass[pra,twocolumn,10pt,aps]{revtex4-1}
\usepackage[utf8]{inputenc}  %Input what you want e.g., é, ł, a, ü
\usepackage[T1]{fontenc}     %Output what you want e.g., é, ł, a, ü
\usepackage[british]{babel}
\usepackage{xcolor}
\usepackage[scaled=0.86]{berasans}  % URL font that go well wtih palatino
\usepackage[colorlinks=true,allcolors=blue]{hyperref}  %Hyperlinks (pink, green, blue)
\usepackage{graphicx} % Package to insert external figures
\usepackage[babel]{microtype}  %Improves text justification
\usepackage{amsmath,amssymb,amsthm,bm,amsfonts,mathrsfs,bbm} %Useful math packages
\usepackage{setspace}
\usepackage{braket}
\usepackage[bottom]{footmisc}
\usepackage{blochsphere}
\usepackage{pgfplots}
\usetikzlibrary{calc,3d,shapes, pgfplots.external, intersections}

\usepackage[bitstream-charter]{mathdesign}
\usepackage{tikz}
\usepackage{mathtools}
\usetikzlibrary{arrows.meta}
\usepackage{csquotes}

\usepackage{xspace}  %Useful to add space in macros
\usepackage{pgfplots}
\usepackage{verbatim}

\newcommand\id{\leavevmode\hbox{\small1\kern-3.3pt\normalsize1}}

\begin{document}

%\preprint{APS/123-QED}

\title{Broadcasting of NPT Entanglement in Two Qutrit Systems}
\author{ Rounak Mundra$^{1}$, Sabuj Chattopadhyay$^{1}$,  Indranil Chakrabarty$^{2}$, Nirman Ganguly$^{3}$}
\email[rm29031996@gmail.com\\
deepsabuj@gmail.com\\
 indranil.chakrabarty@iiit.ac.in\\
 nirmanganguly@hyderabad.bits-pilani.ac.in]{}
\affiliation{$^{1}$Center for Computational Natural Sciences and Bioinformatics, International Institute of Information Technology-Hyderabad, Gachibowli, Telangana-500032, India.\\
$^{2}$Center for Security, Theory and Algorithmic Research, International Institute of Information Technology-Hyderabad, Gachibowli, Telangana-500032, India.\\
$^{3}$ Department of Mathematics, Birla Institute of Technology and Science Pilani, Hyderabad Campus, Telangana-500078, India. \\
}

\begin{abstract} It is known that beyond $2 \otimes 2$ and $2 \otimes 3$ dimensional quantum systems,  Peres-Hordecki criterion is no longer sufficient as an entanglement detection criterion as there are entangled states with both positive and negative partial transpose (PPT and NPT). Further, it is also true that all PPT entangled states are bound entangled states. However, in the class of NPT states, there can exist bound entangled states as well as free entangled states. All free/useful/distillable  entanglement is a part of the class of NPT entangled states. In this article,  we ask the question that given an NPT entangled state in $3 \otimes3$ dimensional system as a resource, how much entanglement can we broadcast so that resource still remains NPT. We have chosen $3 \otimes 3$ system as a first step to understand broadcasting of NPT states in higher dimensional systems.
In particular, we find out the range of broadcasting of NPT entanglement for Two parameter Class of States (TPCS) and Isotropic States (IS). Interestingly, as a derivative of this process we are also able to locate the existence of absolute PPT states (ABPPT) in $3 \otimes 3$ dimensional system. Here we  implement the strategy of broadcasting through approximate cloning operations.
%Free entanglement can act as fundamental resources in several important quantum mechanical tasks. Entangled bipartite qutrit states comprises one class of states that provides free entanglement. Therefore generating more of such pairs from a few,  a process known as broadcasting of correlations acquires significance.In this paper, we braodcast correlations by using Heisenberg Cloning Machines. The cloning machine is applied on $3\otimes3$ entangled quantum systems
\end{abstract}

\maketitle

%\tableofcontents

\section{\label{sec:level1}INTRODUCTION}
The last two decades saw the significant rise of quantum technologies  \cite{qtech} such as integrated  quantum  circuits \cite{intqtech} and long haul quantum communication infrastructure and protocols \cite{qcomm} around the world.  These developments have created a notable impact in the field of quantum communication and quantum cryptography (secure communication). Some of the popular protocols in this area of active research include teleportation \cite{teleportation}, secret sharing \cite{secretsharing}, super dense coding \cite{superdensecoding}, key distribution \cite{keydistribution}, digital signatures \cite{digitalsignatures}, and remote entanglement distribution. \cite{remoteentanglementdistribution} It is now possible for many of these protocols to be deployed commercially as well. \cite{intqtech,idquantique}. These technologies not only are far more advantageous than their classical counterparts in terms of their utility and efficiency, but for some of them there exists no such classical analogue. %It is well known that the coherence or superposition of quantum states is primarily responsible for making information processing with quantum systems radically different in its nature from that of classical or digitized information. 
This fundamental difference not only manifests itself in the form of achievements over and above what is feasible from classical information processing, but also in the form of restrictions or constraints on certain tasks.  All such restrictions or impossibilities has the effect of making quantum information, private, secure and permanent. They are generally referred as \enquote{No Go Theorems} of quantum mechanics. \\

Among such No-Go Theorems, \cite{wootters,nogotheorems,bergou,bruss, nobroadcasting,nogotheorems1} the most prominent one is the \enquote{No-Cloning Theorem} \cite{wootters} which prohibits deterministic and noiseless cloning of an arbitrary unknown quantum state \cite{bruss} . However this theorem does not rule out the possibility of approximate \cite{approximatecloning1, cloningbuzek, buzek,buzekoptimal,buzekoptimal1, statedependentmachine} and probabilistic cloning \cite{probcloning}. 
%Approximate cloning was first attempted and achieved through the construction of state independent quantum cloning machines,also known as universal quantum cloning machine (UQCM) and state dependent quantum cloning machine (QCM). 
In approximate quantum cloning machines (AQCM) \cite{approximatecloning1, cloningbuzek, buzek}, the fidelity of cloning is either dependent on the parameters specifying the input state (state dependent quantum cloning machines)\cite{bruss, statedependentmachine} or is fixed and independent of the state parameters (state independent or universal cloning machines) \cite{buzekoptimal,buzekoptimal1}. One such well known example of state independent cloning machine is Buzek-Hillery (B-H) cloning machine, which is \textit{universal} as well as \textit{optimal} with a fidelity $\frac{5}{6}$ \cite{buzekoptimal,buzekoptimal1}. Both of these types of cloning machines are \enquote{symmetric} in the sense that copies at the output port are identical with each other. Beyond symmetric cloners there exist asymmetric cloners having different copies at the output port \cite{optimalassymcloning}.\\

Another impossible operation similar to \enquote{No Cloning} comes from the \enquote{No Broadcasting Theorem} \cite{ nobroadcasting}. This theorem states that it is impossible to broadcast the information present in an unknown quantum state, where \enquote{broadcast} means transmitting the state to multiple recipients. Not only that, we can not even perfectly broadcast quantum correlation and resources in general. However, just as with the case of cloning, even if not perfectly, we can broadcast imperfectly by taking certain modest approaches. This becomes important when there is exigency in creating more number of entangled state in a network which are to be used as resources to facilitate tasks in quantum information processing and distributed computing \cite{kimble2008, wehner2018, guha2019,cirac1999,gottesman2012,broadbent2009}.
One way to accomplish this is by the application of local \cite{buzek-broadcasting, sourav_chat} and non-local (global) \cite{cloningbuzek, sourav_chat} cloning operations. 
%In this technique, two parties namely Alice and Bob, either use local cloners on their individual sub-systems or nonlocal cloner jointly on the entire (combined) system to create two output pairs which remain entangled over a finite range of input state parameters. 
Researchers showed that AQCMs having fidelity over $\frac{1}{2}(1 + \frac{1}{\sqrt{3}})$ can achieve broadcasting of entanglement with local cloning operations. It was also shown that entanglement in the input state is optimally broadcast only when the local quantum cloner  is optimal. Later, it was proved that optimal broadcasting of entanglement is only possible with symmetric cloners  \cite{optimalassymcloning, aditya_jain}. Recent works like \cite{sourav_chat}, \cite {bandyopadhyay}, \cite{aditya_jain} and \cite{udit_sharma} give an exhaustive analysis of  broadcasting of correlation and other resources in $2\otimes2$ dimension by using both symmetric  and asymmetric cloners. Brodcasting of entanglement was also investigated beyond $2\otimes2$ system, especially in $2\otimes d$ system recently \cite{Rounak}. 
A summary of the previous contributions in this direction, in contrast to those in this work is explicitly presented in Table (\hypersetup{linkcolor=green}\ref{table_11}\hypersetup{linkcolor=blue}). \\

\pagebreak
\onecolumngrid
%\begin{widetext}

\begin{table}[ht!]
\setlength{\tabcolsep}{0.3em}
  \begin{center}
    \caption{Summary of earlier results along with those in the present work on broadcasting of entanglement, discord and coherence. The abbreviations such as NME, MEMS, TPCS, IS, 2-qubit general, qubit-qutrit general and qubit-qudit general stand for non-maximally entangled state, maximally entangled mixed state, two parameter class of states, isotropic states, general two qubit mixed state, general qubit-qutrit mixed state, and general qubit-qudit mixed state classes respectively.}
    \vspace{0.2cm}
    \begin{tabular}{c|c|c|c|c}
      \textbf{System's dimension} & \textbf{Resource state} & \textbf{Broadcasting of} & \textbf{Cloning operation} & \textbf{Author(s)}\\
      \hline
      2 $\otimes$ 2 & NME & Entanglement & Symmetric & Buzek \textit{et al.} and Hillery \cite{buzek-broadcasting, cloningbuzek}\\
      
      2 $\otimes$ 2 & NME & Entanglement & Symmetric & Bandyopadhyay \textit{et al.} \cite{bandyopadhyay} \\
      
      2 $\otimes$ 2 & NME & Entanglement & Asymmetric & Ghiu \cite{optimalassymcloning} \\
      
      2 $\otimes$ 2 & 2-qubit general & Entanglement and Discord & Symmetric & Chatterjee \textit{et al.} \cite{sourav_chat}\\
      
      2 $\otimes$ 2 & 2-qubit general & Entanglement and Discord & Asymmetric & Jain \textit{et al.} \cite{aditya_jain} \\
      
      2 $\otimes$ 2 & 2-qubit general & Coherence & Symmetric & Sharma \textit{et al.} \cite{udit_sharma} \\
      
      2 $\otimes$ 3 & qubit-qutrit general & Entanglement & Symmetric & \cite{Rounak} \\
      
      2 $\otimes$ d & qubit-qudit general & Discord, Coherence & Symmetric & \cite{Rounak}\\

      3 $\otimes$ 3 &  TPCS, IS  & Entanglement & Symmetric & This work\\
    \end{tabular}
  \end{center}
  \label{table_11}
\end{table}
%\end{widetext}

\twocolumngrid
In higher dimensions (dimensions above two qubits and qubit-qutrits), it is difficult to ascertain the difference between entangled and separable states, as the PPT (positive partial transpose) criteria is no longer necessary and sufficient. There are entangled states with positive partial transpose. As the first step towards understanding broadcasting of NPT entangled states through cloning operations, here in this article we have chosen $3 \otimes 3$ dimensional system and in particular 1) Two Parameter Class of States (TPCS) and 2) Isotropic States (IS).
However, equally intriguing is the presence of states which retain their PPT property under any global unitary operation. Such states are known as ABPPT (absolute positive partial transpose) states. Global unitary operations are considered to be powerful resources as they can convert a separable state into an entangled state. Therefore, it is quite counterintuitive that ABPPT states exist. Such ABPPT states can be separable or entangled, a distinction which still is an open problem. Constructions of ABPPT states have remained mathematical in current literature. In the present submission we lay down a physical procedure which produces ABPPT states as an output.\\

In section II, we give short introduction to different concepts which will be relevant to the central idea of this article. In section III we take these two class of states we find out the range of broadcasting in terms of the input state parameters such that the output state still remains a NPT entangled state. We also obtain the ABPPT states as a derivative of this process. This vindicates the significance of our protocol. Finally we conclude in section IV.

\section{Definitions And Primary Concepts}
This section introduces several concepts which are going to be used in the article.
\subsection{General qutrit-qutrit mixed state}
We have considered a general qutrit-qutrit mixed entangled state $\rho_{12}$ as a resource state.  This state is canonically represented as:
\begin{equation}\label{general_state}
\begin{split}
\rho_{12} = \frac{1}{9}\Bigg(\mathbb{I}_{3}\otimes\mathbb{I}_{3} + \sum_{i=1}^8 x_i G_i\otimes\mathbb{I}_{3} + \sum_{j=1}^8 y_j\mathbb{I}_{3}\otimes G_j \\
+ \sum_{i=1}^8\sum_{j=1}^8 t_{ij}G_i\otimes G_j \Bigg) = \big\{\vec{X},\vec{Y},T\big\},
\end{split}
\end{equation}
where $x_i = Tr[\rho_{12} (G_{i}\otimes \mathbb{I}_{3})]$, $y_j = Tr[\rho_{12}(\mathbb{I}_{3}\otimes G_j)]$, $T_{ij} =Tr[\rho_{12}(G_{i} \otimes G_{j})]$, $G_j$'s are Gell-Mann matrices and $\vec{X}$, $\vec{Y}$ and $T$ are the Bloch vectors and the correlation matrix respectively. $\mathbb{I}_{3}$ represents identity matrix of order $3 \times 3$.

\subsection{Criterion to detect the entanglement of a quantum system}
The oldest criterion to identify the separability of a bipartite state was given by  Peres-Horodecki (\textbf{PH}) criterion \cite{ ppt_criteria1, ppt_criteria2}. The criterion  states that if the partial transpose  $\rho_{m \mu,\eta v}^{T}$ = $\rho_{m v,\eta \mu}$ of the joint density matrix $\rho$ of the state is negative, then the state is entangled. Equivalently, we can say that if at least one of the eigenvalues of a partially transposed density operator for the state $\rho$ is negative, then the state  is entangled. However this criterion is a necessary criterion in general and  necessary and sufficient only in $2\otimes2$ and $2\otimes3$ dimensional systems. In higher dimensional systems, if partial transposed density operator of the state has a negative eigenvalue, then the state is entangled and is known as the \textbf{NPT entangled state}. However, we cannot conclude when all the eigenvalues are positive. There can be entangled states with positive partial transpose and these states are known as \textbf{PPT entangled state}. So in $3\otimes3$ dimensional system, which is relevant to this article, there can be both PPT as well as NPT entangled states. 

\subsection{Absolute PPT States}
A bipartite quantum state $\rho \in \mathcal{H}_{n}\otimes\mathcal{H}_{n}$ (a Hilbert space of dimension $n\otimes n$) is said to belong to a set of absolutely PPT states if and only if $\mathcal{U}\mathcal{\rho}\mathcal{U^{\dagger}}$ has positive partial transpose (PPT) for all unitary operators $\mathcal{U}\in \mathcal{H}_{n}\otimes\mathcal{H}_{n}$. For the special case of a state $\rho$ $\in \mathcal{H}_{3}\otimes\mathcal{H}_{n}$ to be absolutely PPT, the condition transforms to the eigenvalues ($\mu_{1} \geq \mu_{2} \geq \hdots \geq \mu_{3n}$) of the state satisfying both the semi-definiteness conditions given below \cite{Arun},

\begin{eqnarray}
L_{1}:=\begin{bmatrix}2\mu_{3n}
& \mu_{3n-1}-\mu_{1} & \mu_{3n-3}-\mu_{2}\\ \mu_{3n-1}-\mu_{1}
& 2\mu_{3n-2}& \mu_{3n-4}-\mu_{3}\\ \mu_{3n-3}-\mu_{2}& \mu_{3n-4}-\mu_{3}&2\mu_{3n-5}
\end{bmatrix}\geq0,\\
L_{2}:=\begin{bmatrix}2\mu_{3n}
& \mu_{3n-1}-\mu_{1} & \mu_{3n-3}-\mu_{2}\\ \mu_{3n-1}-\mu_{1}
& 2\mu_{3n-3}& \mu_{3n-4}-\mu_{3}\\ \mu_{3n-2}-\mu_{2}& \mu_{3n-4}-\mu_{3}&2\mu_{3n-5}
\end{bmatrix}\geq0.
\end{eqnarray}

\subsection{Approximate Quantum Cloning}
We already know from the "No Cloning theorem" that for an arbitrary unknown quantum state $|\psi\rangle$, it is impossible to get the two copies of the state $|\psi\rangle$ perfectly. In other words, there does not exist any completely positive trace preserving map  $C$ such that, $ C : |\psi\rangle \rightarrow  |\psi\rangle\otimes|\psi\rangle$, for all $|\psi\rangle $ in the Hilbert space $\mathcal{H}$. However, the theorem never rules out the possibility of approximate cloning. \\

In this work, we have used symmetric version of the optimal universal asymmetric Heisenberg cloning machine. This machine creates the second clone with maximal fidelity for a given fidelity of first one. For the cloning of a generalized qudit, the general unitary transformation by this machine is given by:
\begin{eqnarray}
\begin{aligned}
\label{heisenberg}
U \lvert j\rangle_x\lvert 00\rangle_{yz}  \rightarrow \sqrt{\frac{2}{d+1}}\Big( &\lvert j\rangle_{x}\lvert j\rangle_{y}\lvert j\rangle_{z}\\ +\frac{1}{2}\sum_{r=1}^{d-1}\lvert j\rangle_{x} \lvert \overline{j+r}\rangle_{y}\lvert \overline{j+r}\rangle_{z} 
&+\frac{1}{2}\sum_{r=1}^{d-1}\lvert \overline{j+r}\rangle_{x} \lvert j\rangle_{y}\lvert \overline{j+r}\rangle_{z}\Big),
\label{cloning}
\end{aligned}
\end{eqnarray}
The suffixes '$x$' and '$y$' denote the clones while '$z$' denotes the ancillary state.

\begin{figure}[h]
\begin{center}
\[
\begin{array}{cc}
\includegraphics[height=12.5cm,width=8.5cm]{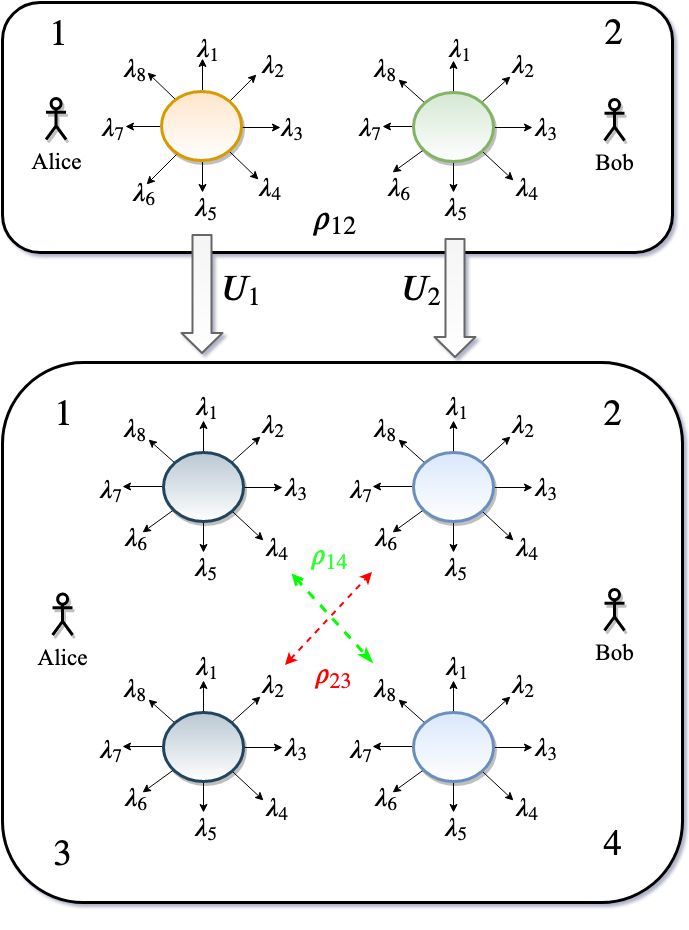}
\end{array}
\]
\end{center}
\caption{\noindent
%\scriptsize
A schematic diagram depicting the application of local cloning unitaries $U_1$ and $U_2$ on a qutrit-qutrit input state $\rho_{12}$ shared between two hypothetical spacelike separated observers named Alice \& Bob to get the non local output states $\rho_{14}$ and $\rho_{23}$. The qutrit system on both sides is illustrated with a sphere having eight arrows ($\lambda_{i}$) which depits the Gell-Mann matrices.}
\label{fig:local_ent} 
\end{figure}

\subsection{Broadcasting of quantum entanglement by cloning}
Quantum entanglement is one the key resources that is required for information processing tasks and distributed computing. Thus the aim of distributing entanglement across various nodes in a network is of practical importance. In a network, there is always a requirement of more entangled pairs. This process of creating larger number of entangled pairs with lesser entanglement from a entangled pair with larger entanglement is termed as "broadcasting of quantum entanglement". In this article, we have used quantum cloning to achieve this task. However there can be several strategies to broadcast entanglement.\\ 

Moving beyond $2 \otimes 2 $ and $2 \otimes 3 $ dimensional systems, the concept of entanglement changes as there can be both \textbf{PPT} and \textbf{NPT entangled states}. In general, PPT entangled states are not useful in information processing tasks whereas NPT entangled states are useful. In this work, we investigate broadcasting of NPT entanglement in a qutrit-qutrit system ($3 \otimes 3 $). Consider two parties Alice and Bob who share a generalized qutrit-qutrit mixed state $\rho_{12}$(\ref{general_state}) as initial input state. We apply local unitary operations for cloning $U_{13}\otimes U_{24}$ on the qutrit pairs $(1,3)$ and $(2,4)$. After carrying out partial trace of the subsystems $(2,4)$ and $(1,3$), we then get the local output states as $\tilde{\rho}_{13}$ and $\tilde{\rho}_{24}$ on Alice's side and Bob's side respectively. Similarly by tracing out the appropriate qutrits, we get two plausible groups of nonlocal output states $\tilde{\rho}_{14}$ and $\tilde{\rho}_{23}$. The representation of the process is provided in Figure \ref{cloning}.\\

The expression for non-local output states across the subsystems of the two spatially separated parties Alice and Bob are expressed by:
\begin{eqnarray}
\begin{aligned}
 \tilde{\rho}_{14}  & = Tr_{23}[\tilde{\rho}_{1234}]\\& = Tr_{23}[U_{13}\otimes U_{24}(\rho_{12}\otimes \rho^{b}_{34}\otimes  \rho^{m}_{56})U_{13}^{\dagger}\otimes U_{24}^{\dagger}],\\
 \tilde{\rho}_{23} & = Tr_{14}[\tilde{\rho}_{1234}] \\
 &= Tr_{14}[U_{13}\otimes U_{24}(\rho_{12}\otimes \rho^{b}_{34}\otimes \rho^{m}_{56})U_{13}^{\dagger}\otimes U_{24}^{\dagger}],\\ 
\end{aligned}
\end{eqnarray}
Additionally, the local output states within their individual subsystems are expressed by:
\begin{eqnarray}
\begin{aligned}
\tilde{\rho}_{13} & = Tr_{24}[\tilde{\rho}_{1234}] \\
& =  Tr_{24}[U_{13}\otimes U_{24}(\rho_{12}\otimes\rho^{b}_{34}\otimes \rho^{m}_{56})U_{13}^{\dagger}\otimes U_{24}^{\dagger}],\\
\tilde{\rho}_{24} & = Tr_{13}[\tilde{\rho}_{1234}] \\
& = Tr_{13}[U_{13}\otimes U_{24}(\rho_{12}\otimes\rho^{b}_{34}\otimes \rho^{m}_{56})U_{13}^{\dagger}\otimes U_{24}^{\dagger}].\\
\end{aligned}
\end{eqnarray}

Here the cloning operations $U_{1}$ and $U_{2}$ are optimal universal asymmetric Heisenberg cloning transformations. The states $ \rho^{b}_{34} = \lvert 00\rangle \langle 00\rvert$ and $ \rho^{m}_{56} = \lvert 00\rangle \langle 00\rvert$ represent the initial blank state and the initial machine state respectively. In order to attain our objective to broadcast \textbf{NPT} entanglement between desired pairs $(1,4)$ and $(2,3)$, we should be able to create \textbf{NPT} entanglement between nonlocal pairs $(1,4)$ and $(2,3)$ irrespective of the local pairs $(1,3)$ and $(2,4)$.

\section{Broadcasting Of Entanglement In $3 \otimes 3$ Dimension}

In this work we mainly focus on two mixed entangled states in $3 \otimes 3$ dimensions, namely : (A) Two Parameter Class of States (TPCS) and (B) Isotropic state. We investigate whether we can broadcast NPT entanglement present in these two states.\\

\subsection{ Two Parameter Class of States (TPCS)}
Consider the following class of states with two real parameters $b$ and $c$ in bipartite qutrit quantum systems : 

\begin{equation}
\begin{split}
\rho_{b,c} = a \sum_{i=0}^{2} \lvert ii\rangle\langle ii\lvert + b \sum_{i,j=0,i<j}^{2} \lvert \psi_{ij}^{-}\rangle\langle \psi_{ij}^{-}\lvert + c \sum_{i,j=0,i<j}^{2} \lvert \psi_{ij}^{+}\rangle\langle \psi_{ij}^{+}\lvert,
\end{split}
\end{equation}
where $\lvert\psi_{ij}^{\pm}\rangle$ = $\frac{1}{\sqrt{2}}(\lvert ij\rangle \pm \lvert ji\rangle)$ and $\{|i\rangle\}  $ represents the vectors in the computational basis.\cite{DiVincenzo} The parameter $a$ is dependent on parameters $b$ and $c$ by unit trace condition, 3 ($a + b + c$) = 1. From unit trace condition, $b$ and $c$ can vary from 0 to $\frac{1}{3}$. However for the matrix $\rho_{b,c}$ to remain positive semi-definite, the parameter '$c$' can take values at most $\frac{1}{3}-b$. This state is NPT entangled when state parameter '$b$' ranges from $\frac{1}{6}$ to $\frac{1}{3}$. It's also worth noting that the states described by points on the line $3b+6c=1$ correspond to qutrit Werner states. \\

This input state $\rho_{12}$ is shared by two parties, Alice and Bob. Both parties apply local cloning machine given in eqn (4) to obtain a composite system $\tilde{\rho}_{1234}$. By tracing out the appropriate qutrits on both sides, we obtain the two plausible groups of non local states as follows :

\begin{eqnarray}
\begin{aligned}
\tilde{\rho}_{14} = \tilde{\rho}_{23} = \bigg\{\vec{X}_{tpcs}, \vec{Y}_{tpcs}, T_{tpcs} \bigg\},
\end{aligned}
\end{eqnarray}
where $\vec{X}_{tpcs} = \vec{Y}_{tpcs} = {0,0,0,0,0,0,0,0}$ and the non zeros entries of the correlation matrix ($T_{tpcs}$) are $t_{1,1}=\frac{25(c-b)}{64}$, $t_{2,2}=\frac{25(c-b)}{64}$, $t_{3,3}=\frac{25(2-9b-9c)}{192}$, $t_{4,4}=\frac{25(c-b)}{64}$, $t_{5,5}=\frac{25(c-b)}{64}$, $t_{6,6}=\frac{25(c-b)}{64}$, $t_{7,7}=\frac{25(c-b)}{64}$, $t_{8,8}=\frac{25(2-9b-9c)}{192}$. Here $t_{i,j}$ denotes the element of $i^{th}$ row and $j^{th}$ column of the correlation matrix. We apply Peres criterion to determine the condition when these nonlocal output state are NPT entangled or not. We observe that it is NPT entangled when state parameter $b$ is greater than $\frac{19}{75}$ and less than $\frac{1}{3}$  which is a subset of the initial parameter range $[1/6,1/3]$.  An explicit example of an NPT entangled TPCS state that can be broadcast and still remain NPT entangled is $\frac{4}{15\sqrt{2}}[(\lvert01\rangle-|10\rangle)(\langle01\lvert-\langle10|)+(\lvert02\rangle-|20\rangle)(\langle02\lvert-\langle20|)+(\lvert12\rangle-|21\rangle)(\langle12\lvert-\langle21|)]+\frac{1}{15\sqrt{2}}[(\lvert01\rangle+|10\rangle)(\langle01\lvert+\langle10|)+(\lvert02\rangle+|20\rangle)(\langle02\lvert+\langle20|)+(\lvert12\rangle+|21\rangle)(\langle12\lvert+\langle21|)]$ with the value of $b=\frac{4}{15}$ and $c=\frac{1}{15}$.\\

To demonstrate the generation of absolute PPT states in non local output states ($\rho_{14}$ and $\rho_{23}$), we generated $10^4$ random values of state parameters $b$ and $c$ from a uniform random distribution. We observed that absolute PPT were found for some state parameters. This is displayed in figure (\ref{fig:local_ent}). The states are represented by brown dots. \\
% The NPT state parameter ranges for which absolute PPT states were found upon broadcasting are given by the following analytical conditions:
% \begin{equation*}
% \begin{split}
% (\frac{1}{6}\leq b\leq\frac{1}{225}\left(-313+2\sqrt{34629}\right)\;\&\&\;0\leq c\leq\frac{1}{3}) || \\ (\frac{1}{225}\left(-313+2\sqrt{34629}\right)<b\leq\frac{1}{3}\;\&\&\;c\geq0\;\&\&\; \\
% 53+\sqrt{3 (97-225b )(767+225b)}) \geq 225 (b+4c))\\
% \end{split}
% \end{equation*}

An example of an NPT entangled TPCS state that can be broadcast to produce ABPPT states is $\frac{2}{15}(\lvert00\rangle \langle00\lvert+\lvert11\rangle \langle11\lvert+\lvert22\rangle \langle22\lvert)+\frac{1}{5}[(\lvert01\rangle-|10\rangle)(\langle01\lvert-\langle10|)+(\lvert02\rangle-|20\rangle)(\langle02\lvert-\langle20|)+(\lvert12\rangle-|21\rangle)(\langle12\lvert-\langle21|)]$ with the value of $b=\frac{1}{5}$ and $c=0$.
\begin{figure}[h]
\begin{center}
\[
\begin{array}{cc}
\includegraphics[scale=0.8,height=8cm,width=8.5cm]{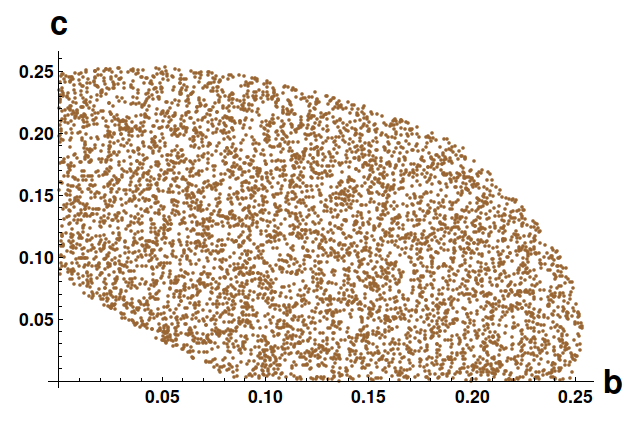}
\end{array}
\]
\end{center}
\caption{\noindent
%\scriptsize
Plot depicting the values (in brown) of two input state parameters: $b$ and $c$ of the state $\rho_{12}$, which will generate the absolute PPT states in $\rho_{14}$ and $\rho_{23}$.}
\label{fig:local_ent} 
\end{figure}

\subsection{ Isotropic States (IS)}
In this subsection, we will consider the class of density matrices, called as isotropic density matrices, which are convex mixture of a maximally entangled state and the maximally mixed state : 
\begin{equation}
\begin{split}
\rho_{f} = \frac{1-f}{d^{2}-1}(\mathbb{I} - \lvert\psi^{+}\rangle\langle\psi^{+}\lvert) + f \lvert\psi^{+}\rangle\langle\psi^{+}\lvert,
\end{split}
\end{equation}

for $0\leq f \leq 1$ and $\lvert \psi^{+}\rangle = \frac{1}{\sqrt{d}} \sum_{i=1}^{d} \lvert ii \rangle$. These states are separable for $f \leq \frac{1}{d}$, and entangled otherwise. For $d = 3$ case, these states are NPT entangled when state parameter '$f$' is greater than $\frac{1}{3}$. Let us assume that this isotropic state is shared by two parties, Alice and Bob. They both apply local cloning transformations given by eqn (4) on their respective qutrits to obtain the composite system $\tilde{\rho}_{1234}$. By tracing out the ancillas and local qutrits on both sides, we obtain the nonlocal output states as follows, 
\begin{eqnarray}
\begin{aligned}
\tilde{\rho}_{14} = \tilde{\rho}_{23} = \bigg\{\vec{X}, \vec{Y}, T \bigg\},
\end{aligned}
\end{eqnarray}

where $\vec{X} = \vec{Y} = \{ 0,0,0,0,0,0,0,0\}$ and the non zero entries of correlation matrix ($T$) are $t_{1,1}=\frac{25(-1+9f)}{768}$, $t_{2,2}=-\frac{25(-1+9f)}{768}$, $t_{3,3}=\frac{25(-1+9f)}{768}$, $t_{4,4}=\frac{25(-1+9f)}{768}$, $t_{5,5}=-\frac{25(-1+9f)}{768}$, $t_{6,6}=\frac{25(-1+9f)}{768}$, $t_{7,7}=-\frac{25(-1+9f)}{768}$, $t_{8,8}=\frac{25(-1+9f)}{768}$. Here $t_{i,j}$ denotes the element of $i^{th}$ row and $j^{th}$ column of the correlation matrix. Now, we again apply Peres criterion to determine the condition when these nonlocal output state are NPT entangled for non-optimal broadcasting. We observe that it is NPT entangled when state parameter $f$ is greater than $\frac{17}{25}$ and less than $1$. An example of an NPT entangled isotropic state which will remain NPT entangled on broadcasting would be $\frac{1}{40}\mathbb{I}+\frac{31}{40}\lvert\psi^{+}\rangle\langle\psi^{+}\lvert$  with the value of $f$ being $\frac{4}{5}$. \\

We also observed that absolute PPT were found in nonlocal output states ($\rho_{14}$ and $\rho_{23}$) when the range of state parameter 'f' is less than $\frac{433}{825}$. An example of an NPT entangled isotropic state that can be broadcast to produce ABPPT states is $\frac{1}{16}\mathbb{I}+\frac{59}{80}\lvert\psi^{+}\rangle\langle\psi^{+}\lvert$  with the value of $f$ being $\frac{1}{2}$.

\section{\label{sec:level1}CONCLUSION} In conclusion we can say that in this work we are able to broadcast NPT entangled states in $3 \otimes 3$ system (TPCS,IS) for certain range of input state parameters. This can be considered a significant step to broadcast NPT entanglement beyond $2 \otimes 2$ and $2 \otimes 3$ systems. Another significant aspect of this submission is that we give a physical procedure as a derivative of which we are able to produce ABPPT states as an output. This work initiates the process of investigating broadcasting of NPT entangled states in higher dimension and broadcasting of other quantum resources like correlation and coherence in $3 \otimes 3$ system. \\

\section*{\label{sec:level1}ACKNOWLEDGEMENT}
 N.G. would like to acknowledge support from the Research Initiation Grant of BITS-Pilani, Hyderabad vide letter no. BITS/GAU/RIG/2019/H0680 dated 22nd April, 2019.
 
% \textit{Acknowledgements:} N.G. would like to acknowledge support from the Research Initiation Grant of BITS-Pilani, Hyderabad vide letter no. BITS/GAU/RIG/2019/H0680 dated 22nd April, 2019.

\end{document}